%% file: 00-main.tex
\documentclass[conference,letterpaper]{IEEEtran}
\IEEEoverridecommandlockouts
\usepackage{cite}
\usepackage{amsmath,amssymb,amsfonts}
\usepackage{algorithmic}
\usepackage{graphicx}
\usepackage{textcomp}
\usepackage{xcolor}
\usepackage{multirow,multicol}
\def\BibTeX{{\rm B\kern-.05em{\sc i\kern-.025em b}\kern-.08em
    T\kern-.1667em\lower.7ex\hbox{E}\kern-.125emX}}
\begin{document}
\title{Toward Smart Scheduling in Tapis\\
\thanks{* This material is based upon work supported by the National Science Foundation Office of Advanced CyberInfrastructure, Collaborative Proposal: Frameworks: Project Tapis: Next Generation Software for Distributed Research (Award \#1931439)}
}

\makeatletter
\newcommand{\linebreakand}{%
  \end{@IEEEauthorhalign}
  \hfill\mbox{}\par
  \mbox{}\hfill\begin{@IEEEauthorhalign}
}
\makeatother
\author{\IEEEauthorblockN{Joe Stubbs}
\IEEEauthorblockA{\textit{Texas Advanced Computing Center} \\
\textit{University of Texas at Austin}\\
Austin, TX, USA \\
jstubbs@tacc.utexas.edu}
\and
\IEEEauthorblockN{Smruti Padhy}
\IEEEauthorblockA{\textit{Texas Advanced Computing Center} \\
\textit{University of Texas at Austin}\\
Austin, TX, USA \\
spadhy@tacc.utexas.edu}
\and
\IEEEauthorblockN{Richard Cardone}
\IEEEauthorblockA{\textit{Texas Advanced Computing Center} \\
\textit{University of Texas at Austin}\\
Austin, TX, USA \\
racardone@tacc.utexas.edu}
}

\maketitle


\begin{abstract}
The Tapis framework provides APIs for automating job execution on remote resources, including HPC clusters and servers running in the cloud. Tapis can simplify the interaction with remote cyberinfrastructure (CI), but the current services require users to specify the exact configuration of a job to run, including the system, queue, node count, and maximum run time, among other attributes. Moreover, the remote resources must be defined and configured in Tapis before a job can be submitted. In this paper, we present our efforts to develop an intelligent job scheduling capability in Tapis, where various attributes about a job configuration can be automatically determined for the user, and computational resources can be dynamically provisioned by Tapis for specific jobs. We develop an overall architecture for such a feature, which suggests a set of core challenges to be solved. Then, we focus on one such specific challenge: predicting queue times for a job on different HPC systems and queues, and we present two sets of results based on machine learning methods. Our first set of results cast the problem as a regression, which can be used to select the best system from a list of existing options. Our second set of results frames the problem as a classification, allowing us to compare the use of an existing system with a dynamically provisioned resource. 
\end{abstract}




\input{01-intro}
\input{03-challenges}
\input{03-methods}

\input{04-results}

\input{02-relatedwork}

\input{05-futurework}

\section*{Acknowledgment}
Supported by NSF awards \#1931439 and \#2112606. Special thanks to Constantinos Skevofilax for data curation.


\bibliographystyle{IEEEtran}
\bibliography{IEEEabrv,smart-sched-references.bib}

\end{document}

%% file: 01-intro.tex
\section{Introduction}
Tapis \cite{tapisRefJstubbs2021} is a cloud-hosted API framework for reproducible computational research with thousands of active users. A primary feature of Tapis is the ability to execute jobs on remote systems on behalf of users, including batch jobs on HPC clusters and high-throughput jobs on cloud servers. While an individual user may have access to several systems -- from multiple HPC clusters at different centers, each with different queues, to cloud and high-throughput servers running on campus clusters or public clouds -- Tapis currently requires the user to specify the exact parameters for each job, including the system to run on, the queue, if applicable, and other aspects, such as the number of nodes to use, the maximum runtime for the job, etc. Moreover, the system resource to be used for the job must be configured in Tapis prior to submitting the job, precluding solutions where resources are provisioned automatically and "just in time" for a specific user's workload. While in some cases users may wish to specify exact details about where and how their job should run, in other cases, the user's primary objective is simply to get the analysis performed as quickly as possible. Additionally, they may lack detailed knowledge about the characteristics of different systems, queues and/or the application, making it difficult or impossible for them to provide an optimal job configuration.

In this paper, we present our work to date to develop a smart scheduling job capability in the Tapis framework. We define the \textit{smart scheduling with dynamic resource provisioning} problem as follows: given a user-supplied job submission request with partial configuration, automatically determine the complete job configuration which optimizes some objective function, considering all systems/queues available to the user and, if applicable, the possibility of dynamically provisioning dedicated resources for the job. In general, the objective function can be thought of as a cost function to be minimized, where cost could be measured in different ways, e.g., time, service units, dollars, CO2 emissions, etc. In this work, we focus on the objective of minimizing \textit{time-to-solution}, that is, the total time from when the user submits the job to when the results of the job are available. 

Developing such a feature presents a number of challenges which we describe in detail in Section \ref{sect:challenges}, including automatically determining the following: 1) \textit{system attributes}, including hardware architecture and software dependencies; 2) \textit{job constraints}, including required hardware resources (CPU cores, memory) and total runtime; 3) \textit{data movement cost}, that is, the time required to stage input data into and archive resulting output from the execution host, and 4) \textit{system provisioning and queue time}, that is, the total time to provision a resource or wait for a job (in queue) to start running. We believe these challenges represent core problems to be tackled by the research CI community, the solutions to which would enable improved usability and utilization of the underlying cyberinfrastructure. 

In the remaining sections of the paper, we focus on challenge 4) and machine learning (ML) methods for estimating queue time for batch-scheduled HPC systems. Using historical data from the Stampede2 system at the Texas Advanced Computing Center (TACC), we develop two sets of ML models for predicting queue time. In the first case, we develop regression models that predict the real-valued queue time for a job based on a set of six attributes. Two attributes (\textit{num\_nodes} and \textit{max\_minutes}) are part of the job resource request while the remaining four (\textit{backlog\_minutes}, \textit{backlog\_num\_jobs}, \textit{running\_num\_jobs}, and \textit{running\_minutes}) capture current queue state at the time of job submission. These predictions can be used to compare queue time across a set of existing systems and queues. In the second set of results, we use the same six attributes to develop classification models that predict the \textit{queue time bin}, that is, the range of minutes that the job's queue time will likely fall into. The predicted queue time bins organize jobs according to the model's confidence in their candidacy for dynamic provisioning, where jobs classified in the highest bins represent candidates for which the system has the greatest confidence that the dynamic provisioning cost will be considered worth paying. 

We utilize a software framework for automatically searching across a configurable space of model types and hyperparameters. Our best classification and regression models use Histogram-based Gradient Boosting and achieve over 90\% accuracy (respectively, over 0.9 $R^2$ score) on held-out test sets across a six month range of time. These and related results are presented in Section \ref{sect:results}. 





%% file: 03-challenges.tex
\section{Smart Scheduling and Associated Challenges}
\label{sect:challenges}
Our goal is to minimize \textit{time-to-solution} by predicting queue time. Let \textit{time-to-solution} = \textit{initiation\_time} + \textit{data\_move\_time} + \textit{execution\_time}, where \textit{initiation\_time} is either the time spent in queue or the time spent provisioning infrastructure. This paper examines only the \textit{initiation\_time} component of \textit{time-to-solution}.   

In drilling down into the problem of dynamically scheduling batch jobs, we focus on Tapis's design but the challenges are broadly applicable. Our first task is to select the set of candidate hosts on which a job can run.  Each system is characterized by its hardware and software attributes, and each application is constrained to run only on systems with the attributes it requires. For example, an application may be constrained to run only on x86-64 processors, so hosts with ARM processors are excluded. Applications and systems must be compatible: Systems advertise their attributes, and applications are matched to systems that meet their constraints.

\subsection{System Attributes and Job Constraints}
Standards such as Redfish \cite{redfish} and SNMP \cite{snmp} provide extensive vocabularies of system traits.  These vocabularies fully characterize a host's installed hardware, firmware and software.  However, they also introduce a level of complexity and precision not necessary at the application level.

In 2021, SGCI published a resource description specification \cite{sgcipaper, sgcispec} to improve interoperability between data centers and, specifically, to allow applications to select hosts on-the-fly based on compatibility and available capacity. Integration with several frameworks including Tapis was prototyped, but promoting a new standard proved to be resource intensive.

Tapis's current design direction is to use dynamic sets of system attributes that do not have to be predefined.  System definitions can be annotated with key/value pairs to advertise their capabilities. Applications specify boolean expressions that reference system attributes and are evaluated at runtime.  When an expression returns true, the system meets the application's constraints and becomes a candidate for execution. This dynamic approach is flexible, easy to implement and requires only local agreement on vocabulary among applications and the systems they use.     

\subsection{System provision and queue time}
While high-throughput systems such as cloud servers typically start user workloads immediately, most HPC systems utilize a batch scheduler where submitted jobs wait in queue to execute. The wait times vary and depend on factors such as the current state of the queue as well as characteristics of the job. The queue time can represent a significant portion of the overall time to solution; some jobs at TACC can wait in queue for more than 24 hours before starting. Thus, estimating the queue time for a specific job and resource represents an important aspect of computing the time-to-solution objective function.

Dynamically provisioning a resource for a specific job broadly involves instantiating virtual servers with the storage, networking and software required for the application to run correctly using an API such as AWS EC2 or JetStream2's OpenStack API. Various methods exist for minimizing the total time required for resource provisioning. In this paper, we assume a fixed (i.e., constant) provisioning time to dynamically deploy a job-specific resource. Of course, dynamically provisioning a resource still consumes physical resources from some system and has the potential for other impacts, such as incurring a real cost (in dollars) when using resources on a public cloud. In practice, the consideration to use a dynamically provisioned resource for a job could be a complex decision depending on various factors, such as time-sensitivity of the computation and the availability of other resources. We formalize this trade-off by introducing a \textit{tolerance factor}, a positive real value quantifying the extent to which using existing systems is preferred over dynamically provisioning. To be precise, let $q$ denote the time a job waits in queue on an existing system, let $p$ denote the time to dynamically provision a resource for a job, and let $t$ be the tolerance factor. Then dynamically provisioning a job is desirable whenever $q > p*t$, while $q \leq p*t$ implies that waiting in queue on the existing system is desirable.

In the rest of the paper, we focus on predicting queue times for jobs on existing clusters with the goal of comparing candidate systems to each other as well as to a dynamically provisioned resource. For comparing existing systems to each other, we frame the problem as a regression where the system with the smallest predicted queue time would be selected. For comparing existing systems to a dynamically provisioned resource, we analyse it as a classification problem, where the goal is to predict the queue time bin of a job on a system. Queue time bins are continuous ranges of minutes, and we study the classification problem for different numbers of bins of size $p*t$. In this setting, jobs predicted to be in bin two or higher are candidates for dynamic provisioning, with jobs in the highest bin being the candidates predicted to benefit the most.



%% file: 03-methods.tex
\section{Methods}
\subsection{Data Sources and Preprocessing}

The Texas Advanced Computing Center (TACC) uses tacc\_stats \cite{taccstats} and Slurm to record information about every job submitted to any of its HPC systems, which gives us queue time in minutes for each job.  We focused our attention on 6 months of cleaned, historical data for Stampede2. Specifically, we worked with 2022 data (Feb 1 to Jul 31) for two production queues, \textit{skx-normal} and \textit{normal}, which schedule jobs on SKX and KNL nodes, respectively. 

\subsection{Exploring different techniques }
\label{subsec:tech}
\subsubsection{General approach} 
\label{subsubsec:ga}
We performed standard exploratory data analysis, visualization, and feature selection using the Python \texttt{pandas} and \texttt{matplotlib} packages. We removed features such as \textit{jobid}, \textit{user}, \textit{start\_time} and \textit{end\_time} from the training, test, and validation sets because they were either non-predictive in early results or required future knowledge. We selected the features \textit{num\_nodes} and \textit{max\_minutes}.  We then engineered features that reconstruct system state \textit{at the time each job was submitted}. Specifically, we derived for each job \textit{backlog\_num\_jobs} and \textit{backlog\_minutes} to record the number of jobs in queue and their total requested minutes, respectively. Similarly, \textit{running\_num\_jobs} and \textit{running\_minutes} represent the number of running jobs and their total requested minutes. We applied standard techniques such as data shuffling and handled outliers. Any job whose waiting time was greater than two days was considered an outlier, given the maximum duration for a job request in Stampede2 is two days. This resulted in the removal of 5831 jobs (2.26\%) from the \textit{skx-normal} queue dataset (169114 jobs) and 1821 jobs (1.07\%) from the \textit{normal} queue dataset (257053 jobs).

We developed a model search and evaluation framework in Python that utilizes a configuration file to explore various model types and associated hyperparameter spaces, as well as other configurations such as the dataset to use. For a given input dataset, we developed a configurable sliding window method to split sub-intervals of the data into a \textit{current} and \textit{future} set. The \textit{current} dataset was further divided into training and test sets using an 80/20 split. For example, given an initial data set with six months of jobs data, we could create six one-month windows and split each window into a \textit{current} and \textit{future} set. We experimented with window splitting based on time as well as job counts. Models would then be trained on the training subset of the \textit{current} set and evaluated on the test subset of \textit{current} as well as the \textit{future} set. All of these settings can be assigned in the configuration file, allowing the program to run for days uninterrupted.  


\subsubsection{Regression techniques} 
\label{subsubsec:rt}
Queue time prediction can be modeled as a regression problem. Several regression techniques have been used in the literature based on different use cases and data \cite{menearhpcruntime2023, brown22}. We modeled the problem both as a time-series and non-time-series regression. For non-time-series models, we selected the six features listed in Section \ref{subsubsec:ga} and applied the following machine learning techniques: Linear Regression, K-Nearest Neighbor (kNN), and Histogram-based Gradient Boosting (HGB). For time-series models, we added extra features commonly used in time-series data preparation. We then applied Linear Regression, Lasso, Feed-forward Neural Network, and Long Short-Term Memory by partitioning the data and sliding the window to train and test the data. In the end, we observed that the results were worse than in the non-time series cases.  

\subsubsection{Classification techniques} 
\label{subsubsec:ct}
We also modeled queue time as a classification problem, where the goal is to predict the correct queue time bin for each job. We define queue time bins as continuous ranges of minutes of a fixed size, except for the last bin, which has no upper bound. We explored different sizes and number of bins. For example, in the case of 4 bins of size 60 minutes, the goal is to predict whether the job's queue time will be: 0-60 minutes, 60-120 minutes, 120-180 minutes or greater than 180 minutes. As in the case of regression, we explored various models and associated hyperparameter spaces, including Histogram-based Gradient Boosting (HGB), K-Nearest Neighbor (kNN), Logistic Regression, Random Forrest, and Support Vector Machines. We also used the sliding window technique for training and testing, using accuracy against known queue times as our validation metric.


%% file: 04-results.tex
\section{Evaluation and Results}
\label{sect:results}

We settled on a 90/10 \textit{current/future} split in each of six 1 month windows as described above, and further partitioned \textit{current} into an 80/20 \textit{training/test} split. We modeled using both regression and classification with the features described in Section \ref{subsubsec:ga}. We tried the seven regression and five classification techniques listed in Sections \ref{subsubsec:rt} and \ref{subsubsec:ct}, respectively.

Table \ref{tab:eval} shows the kNN and HGB outcomes, which produced our best results. We highlight only HBG results here because of its computational advantage. We used three metrics for all regression techniques to evaluate the models: $R^2$ score (r2Score), mean absolute error (MAE), and root mean square error (RMSE), the latter two in minutes. We ran HGB regressor from scikit-learn \cite{scikit-learn} with maximum trees = 500 and maximum tree depth = 9. For \textit{skx-normal}, it yielded a high accuracy of r2\_score=0.9 and low MAE=57.85 and RMSE=147.72. These predictions could be used to dynamically choose a system and queue to run a job.

We ran HGB Classifier to predict the bin into which a job's queue time will fall. For \textit{normal}, the accuracy score was a high 0.95 with a last bin rescheduling accuracy of 0.92. Using 60 minutes bins, the last bin contains jobs with a predicted queue time greater than 4 hours, which makes them the best candidates for dynamic provisioning.

In conclusion, the HGB models developed in the study obtained a sufficiently high $r2$ score ($0.9$, regression) and accuracy ($0.92$, classification) to be suitable for use in a smart scheduling application. 

\begin{table*}[]
\centering
\caption{Evaluation Results}
\label{tab:eval}
\begin{tabular}{|l|l|l|l|l|l|l|l|l|l|l|}
\hline
\multirow{2}{*}{Queue} & \multirow{2}{1.5cm}{\# Jobs (Feb-Jul'22)} & \multirow{2}{*}{\#Days/W}& \multirow{2}{*}{\#Jobs/W} & \multicolumn{4}{l|}{Regression} & \multicolumn{3}{l|}{Classification}\\ \cline{5-11} 
  &  & & & Alg.& r2Score & MAE & RMSE & Alg. & Acc. & LastbinAcc \\ \hline
\multirow{2}{*}{skx-normal} &\multirow{2}{*}{169114} & 30& 33822 & HGB & 0.9 & 57.85 & 147.72 & HGB & 0.92 & 0.91  \\ \cline{5-11} 
 & &  &  & kNN & 0.88 & 44.46 & 152.21 & kNN & 0.95 & 0.93\\ \hline

\multirow{2}{*}{normal} &\multirow{2}{*}{257053} & 30 & 51410 & HGB & 0.83 & 30.7 & 87.86 & HGB &0.95 &0.92 \\ \cline{5-11} 
 & &  &  & kNN & 0.83 &  16.66&83.166 & kNN & 0.95& 0.9355\\ \hline
\end{tabular}
\end{table*}

%% file: 02-relatedwork.tex
\section{Related Work}
Several works have used machine learning approaches to predict queue waiting time and job runtime using HPC job historical data \cite{brown22, menearhpcruntime2023}. 
In \cite{brown22}, the authors used machine learning approaches to predict job start time quickly and accurately to help place urgent workloads across HPC machines. They proposed a stochastic method to generate random queue states that capture the machine usage patterns and use that as input for the model. They used a combination of boosted trees classification and regression models from the XGBoost library with the proposed stochastic method, which improved the accuracy significantly. They can accurately predict the job's start time 85\% of the time within 60 minutes,   90\%  of the time within 2 hrs, and 95\%  of the time within 6 hrs on all three HPC machines' standard queues considered in the paper. Their results are comparable with ours, where the last bin rescheduling accuracy, i.e., jobs with 4 hrs or more queue time, is 91\%.
In \cite{menearhpcruntime2023}, the authors surveyed the runtime prediction studies and recommended a systematic approach to developing a machine learning runtime prediction model. They evaluated their approach on the NREL Eagle HPC dataset. They found that the XGBoost model performs well with a training window of 100 days and a testing window of one day, indicating that the job runtime prediction model needs to be retrained daily. In our study, we found Histogram-based Gradient Boosting regressor and classifier models training on 22 days, testing on 5 days, and validating on future 3 days data give sufficient accuracy for the smart-scheduling use case.

%% file: 05-futurework.tex
\section{Conclusion and Future Work}
We presented our efforts and challenges toward developing smart job scheduling in Tapis. We formulated the problem to minimize the time-to-solution by predicting the job queue wait time. We applied
machine learning techniques for the prediction and observed that histogram-based gradient boosting methods gave us high accuracy on TACC's Stampede2 data, making it a good candidate to leverage in Tapis's smart scheduling design.
In the future, we would like to fine tune our models using different HPC machines' data. Given recent advancements in using large language models (LLMs) on tabular data, we would also like to explore their applicability to our data.